\documentclass[10pt,romanappendices,journal,twocolumn]{IEEEtran}
\newcommand{\fsize}{0.48}

\usepackage{amsmath, amsfonts, amssymb,epsfig, epstopdf, color,amsthm}
\usepackage{citesort}
\usepackage[noadjust]{cite}

\newtheorem{lemma}{Lemma}

\newtheorem{proposition}{Proposition}

\begin{document}
%
\title{Space-Constrained Massive MIMO: Hitting the Wall of Favorable Propagation}
\author{Christos~Masouros,~\IEEEmembership{Senior Member,~IEEE}, and Michail~Matthaiou,~\IEEEmembership{Senior Member,~IEEE}  
%
\vspace{-15pt}



\thanks{Manuscript received February 2, 2015. The associate editor coordinating the review of this letter and approving it for publication was M. Elkashlan.}
 \thanks{%
 C. Masouros is with the Dept. of Electrical \& Electronic Eng., University College London, Torrington Place, London, WC1E 7JE, U.K. (e-mail: chris.masouros@ieee.org)}
 \thanks{%
 M. Matthaiou is with the School of Electronics, Electrical Engineering
and Computer Science, Queen's University Belfast, Belfast, BT3 9DT, U.K.,
and with the Department of Signals and Systems, Chalmers University of
Technology, 412 96 Gothenburg, Sweden (e-mail: m.matthaiou@qub.ac.uk).}
 \thanks{%
 The work was supported by the Royal Academy of Engineering, UK and the Engineering and Physical Sciences Research Council (EPSRC) project EP/M014150/1.}
 }


\maketitle
\begin{abstract}
The recent development of the massive multiple-input multiple-output (MIMO) paradigm, has been extensively based on the pursuit
of \textit{favorable propagation}: in the asymptotic limit, the channel vectors become nearly orthogonal and inter-user interference tends to zero \cite{mMIMOLarsson}. In this context, previous studies have considered fixed inter-antenna distance, which implies an increasing array aperture as the number of elements increases. Here, we focus on a practical, space-constrained topology, where an increase in the number of antenna elements in a fixed total space imposes an inversely proportional decrease in the inter-antenna distance. Our analysis shows that, contrary to existing studies, inter-user interference does not vanish in the massive MIMO regime, thereby creating a saturation effect on the achievable rate.

 \end{abstract}

\IEEEpeerreviewmaketitle

\vspace{-4pt}
\section{Introduction}
Massive multiple-input-multiple-output (MIMO) systems with hundreds of antennas are envisaged for the next generation of base stations (BSs) to meet the ever increasing quality of service (QoS) demands in a power efficient manner \cite{mMIMOLarsson}-\cite{mMIMO}. Existing studies have shown that a key characteristic of massive MIMO is that the inner product of two distinct channel vectors between the BS and the mobile users tends to zero as the number of BS antennas increases \cite{mMIMOLarsson}, \cite{favorablePropagation}. In other words, the multi-user spatial streams become asymptotically pairwise orthogonal. Under this condition, coined \textit{favorable propagation}, low-complexity precoding/detection (e.g. maximum-ratio-transmission (MRT)/maximum-ratio-combining (MRC)), can achieve near-optimal performance \cite{mMIMOLarsson}. 

A major obstacle towards increasing the numbers of antennas in practical systems is the limited available physical space in both BSs and mobile devices. The placement of antenna elements at more than half a wavelength apart is, in general, considered to secure minimal correlation between the communication channels, allowing the exploitation of the full system spatial diversity. On the other hand, a dense deployment of antenna elements increases the spatial correlation, thereby deteriorating the system performance \cite{mMIMO_fixedSpace,mMIMO_fixedSpace2}. Accordingly, for a fixed total physical space, this introduces a tradeoff in the spatial diversity. With an increasing number of antenna elements, along with the increasing signal sources, spatial correlation increases as well, thereby inducing a saturation in the spatial diversity benefits. This observation motivates a study on whether the favorable propagation characteristics hold for the case of space-constrained massive arrays.

In line with the above, in this paper we study the statistical properties of the inner product of channel vectors
for a system equipped with a massive uniform linear array (ULA) of fixed total space, where the inter-antenna distance is inversely proportional to the number of elements and the antenna correlation is directly dependent on the number of antennas. In contrast to \cite{mMIMO_fixedSpace,mMIMO_fixedSpace2}, where the sum rates were derived for particular precoding schemes, here we elaborate on the fundamental statistical properties of the massive MIMO channel for a space-constrained antenna deployment. \color{black} Our analysis shows that the inner product of two distinct channel vectors, which intimately represents inter-user interference, converges to a non-zero value; this causes a saturation of the achievable sum rates under MRT with an increasing numbers of antennas. In fact, it is analytically shown that the first and second-order moments of this inner product converge to non-vanishing values which depend on the total physical space of the BS array. Note that the findings under this scenario stand in fundamental contrast with those of \cite{mMIMOLarsson,favorablePropagation, favorablePropagation2}, which demonstrated that inter-user interference converges smoothly to zero for physically unconstrained antenna arrays.


\vspace{-5pt}
\section{System Model and Achievable Sum Rate}
\vspace{-1pt}
\subsection{Channel Model}
Consider a downlink multiuser MIMO system with a $N$-antenna BS and $K$ single antenna receivers with $N\geq K$. The physical space for the antenna array at the transmitter is constrained to $d_0\lambda$, where $\lambda$ is the carrier wavelength, and $d_0$ is the arbitrary physical constraint in units of wavelengths. 
The channel is assumed to experience frequency flat fading, such that the received signal at the $k$-th terminal reads as
\vspace{-5pt}
\begin{equation} \label{rx} 
y_k=\sqrt{\rho}\mathbf{h}_k^H \mathbf{w}_k s_k+ \sqrt{\rho} \sum_{j=1,j\neq k}^K\mathbf{h}_k^H \mathbf{w}_j s_j + n_k
\end{equation} 
where $(\cdot)^H$ denotes Hermitian transpose, $n_k \sim \mathcal{CN} (0,1)$ is the additive white Gaussian noise vector, $\rho$ is the average signal to noise ratio (SNR), $\mathbf{h}_k^H$ represents the downlink channel to the $k$-th terminal, $\mathbf{w}_k$ and $s_k $ are the unit-norm beamforming vector and the information symbol from the BS for the $k$-th terminal, respectively. For a fixed total space of $d_0$ at the BS, the inter-antenna separation is given as\footnote{Note that, depending on the specific ULA configuration, the total space may be split in $N-1$ equal separations, giving $d=\frac{d_0}{N-1}\lambda$. For notational convenience we herein consider $N$ such separations, assuming $d/2$ additional space at each end of the array, which for large $N$ will give a negligible difference between the two models.}

\begin{equation} \label{d} 
d=\frac{d_0}{N}\lambda.
\end{equation} 
The generic channel model using steering vector representation can be expressed as
\begin{equation} \label{channel}
\mathbf{h}_k=\beta_{k,0}\mathbf{g}(\theta_{k,0})+\sum_{\ell=1}^L\beta_{k,\ell}\mathbf{g}(\theta_{k,\ell})
\end{equation}

\noindent where the first term models the line-of-sight (LOS) component with path gain $\beta_{k,0}$ and the second term accounts for the $L$ multipath components with path gains $\beta_{k,\ell}, \ell \in [1,L]$. The steering vectors $\mathbf{g}_k$ are modeled as
\begin{equation} \label{sv}
\mathbf{g}(\theta_{k,\ell})=\left[1, e^{-i2\pi \frac{d}{\lambda}\sin\theta_{k,\ell}} , ..., e^{-i2\pi\left(N-1\right)\frac{d}{\lambda}\sin\theta_{k,\ell}}\right]^T
\end{equation}
where $(\cdot)^T$ denotes the matrix transpose, $i\triangleq\sqrt{-1}$ and $\theta_k$ is the angle of departure (AoD) for the $k$-th terminal. In line with \cite{favorablePropagation}, \cite{favorablePropagation2} we focus on the uniform random LOS channel with $\beta_{k,0}=1$ and $\beta_{k,\ell}=0$ for all $\ell \neq 0$, for which we have
\begin{equation} \label{LOS}
\mathbf{h}_k=\left[1, e^{-i2\pi \frac{d_0}{N}\sin\theta_k} , ..., e^{-i2\pi\left(N-1\right)\frac{d_0}{N}\sin\theta_k}\right]^T.
\end{equation}

\noindent It can be seen that now the channel response and the resulting antenna correlation are a function of the number of antennas $N$. We assume that all terminals are randomly distributed within a circle-shaped cell with radius $R$. Following \cite{favorablePropagation, favorablePropagation2}, $\sin\theta_k, k=1,...,K$ are assumed to be uniformly distributed within the interval $[-1,1]$.

\subsection{Achievable Sum Rate}
To keep our analysis insightful, we assume that both the BS and users have perfect channel
state information. According to \eqref{rx}, the achievable ergodic rate
of the $k$-th user can be expressed as
\begin{equation} \label{rate}
R_k = E \left\{\log_2 (1 + \textrm{SINR}_k)\right\} 
\end{equation}
where $E\{\cdot\}$ denotes the expectation operator and 

\begin{equation} \label{SINR}
\textrm{SINR}_k \triangleq \frac{\rho|\mathbf{h}_k^H \mathbf{w}_k|^2}{1+\rho\sum_{j=1, j \neq k}^K|\mathbf{h}_k^H \mathbf{w}_j|^2}.
\end{equation}

Considering MRT where the precoding vector is given as $\mathbf{w}_k =\mathbf{h}_k/\sqrt{N}$, the
achievable ergodic rate of the $k$-th user can be rewritten as
\begin{align} \label{rate2}
R_k &= E \left\{\log_2 \left(1 + \frac{\rho N}{1+\frac{\rho}{N}\sum_{j=1, j \neq k}^K|\mathbf{h}_k^H \mathbf{h}_j|^2}\right)\right\}\\
&\geq \log_2 \left(1 + \frac{1}{\frac{1}{\rho N}+(K-1)E \left\{\frac{1}{N^2}|\mathbf{h}_k^H \mathbf{h}_j|^2\right\}}\right)\triangleq R_L \label{rate3}
\end{align}
where \eqref{rate3} has been obtained by applying Jensen's inequality on the convex function of the form $\log_2 \left(1 + \frac{1}{x}\right)$.  The above expression illustrates that the inner product, $\mathbf{h}_k^H \mathbf{h}_j$, between two distinct channel vectors  determines the resulting inter-user interference and, consequently, the achievable rate of each terminal. In the following, we will investigate the statistical properties
of this inner product and show that, in the asymptotic limit as $N \rightarrow \infty$, its first two moments converge to non-zero values.
%

\section{Non-Favorable Propagation Conditions}

In this section, we derive exact and asymptotic expressions for the first and second-order moments of
the inner product $\mathbf{h}_k^H \mathbf{h}_j$, which determine the inter-user interference power. 
We begin with the exact analysis, which holds for any finite $N$, and present the following proposition:
\begin{proposition} For physically constrained ULAs with a total space of $d_0$ in the LOS channel, the first two moments of the inner product 
$\mathbf{h}_k^H \mathbf{h}_j$ are equal to
\begin{align} 
E \Big\{\mathbf{h}_k^H \mathbf{h}_j\Big\}  =& \sum_{m=0}^{N-1} {\rm sinc}^2\left(am\right)\label{exact_mom1}\\
{\rm var} \Big\{\mathbf{h}_k^H \mathbf{h}_j\Big\} =& \sum_{m_1=0}^{N-1}\sum_{m_2=0}^{N-1}{\rm sinc}^2\left(a(m_1-m_2)\right)\notag\\
&-\left(\sum_{m=0}^{N-1} {\rm sinc}^2\left(am\right)\right)^2\label{exact_mom2}
\end{align}
where $a\triangleq 2\pi\frac{d_0}{N}$ for brevity and ${\rm var} (\cdot)$ returns the variance of a random variable.
\end{proposition}

\begin{IEEEproof}
Let us define $u_k\triangleq\sin \theta_k$. Then, we have
\begin{align} \label{I}
E \Big\{\mathbf{h}_k^H \mathbf{h}_j\Big\}= \sum_{m=0}^{N-1}E \left\{e^{i2\pi\frac{d_0}{N}m(u_k-u_j)}\right\}
\end{align}
from which, for a uniformly distributed $u_k$ in the interval $[-1,1]$, it can be seen that 
\begin{align*} 
E \left\{e^{i2\pi\frac{d_0}{N}mu_k}\right\}=\frac{1}{2}\int_{-1}^{1}e^{i2\pi\frac{d_0}{N}mx}dx=\frac{\sin(2\pi\frac{d_0}{N}m)}{2\pi\frac{d_0}{N}m}. 
\end{align*}
The desired results in \eqref{exact_mom1}-\eqref{exact_mom2} follow trivially after some basic algebraic manipulations
and noting that
\begin{equation} \label{Var_def}
{\rm var} \Big\{\mathbf{h}_k^H \mathbf{h}_j\Big\}  = E \left\{\left|\mathbf{h}_k^H \mathbf{h}_j \right| ^2\right\}
-\Big|E \left\{\mathbf{h}_k^H \mathbf{h}_j  \right\}\Big|^2.
\end{equation}
\end{IEEEproof}

We now turn our attention to the massive MIMO regime, by letting $N \rightarrow \infty$. Under these conditions, we obtain
some very important insights. 
\begin{proposition}\label{Asy_prop}
 For physically constrained ULAs with a total space of $d_0$ in the LOS channel, the first two scaled moments of the inner product 
$\mathbf{h}_k^H \mathbf{h}_j$ and the square scaled inner product $|\mathbf{h}_k^H \mathbf{h}_j|^2$ \color{black} tend asymptotically ($N \rightarrow \infty$)  to
{{\begin{align} 
E \bigg\{\frac{1}{N}\mathbf{h}_k^H \mathbf{h}_j\bigg\} & \rightarrow \frac{1}{2N}+\frac{1}{4d_0}\label{asy_mom1}\\
  E \left\{\frac{1}{N^2}|\mathbf{h}_k^H \mathbf{h}_j|^2\right\} & \rightarrow \frac{1}{2d_0}-\epsilon\label{asy_sqp}\\ 
 {\rm var} \bigg\{\frac{1}{N}\mathbf{h}_k^H \mathbf{h}_j\bigg\} & \rightarrow \frac{1}{2d_0}\left({1}-\frac{1}{2N} -\frac{1}{8d_0} \right)-\frac{1}{4N^2}-\epsilon\label{asy_mom2}
\end{align}}}
where $\epsilon=\frac{1}{4\pi^2d_0^2}\sum_{m=1}^{\infty}\frac{\sin^2(am)}{m}$ is a correction term.
\end{proposition}
\begin{IEEEproof}
See Appendix \ref{app:LARGE}.
\end{IEEEproof}

The above results indicate that the channel vectors do not become pairwise orthogonal in the massive MIMO limit.
Instead, their mean inner product converges to a non-zero value that is inversely proportional to the total physical space of the array.
In other words, the smaller the physical length of the BS array, the more we deviate from favorable propagation conditions. 
Yet, if $d_0$ grows analogously with $N$, the mean inner product will converge smoothly to zero, as predicted in 
\cite{mMIMOLarsson,favorablePropagation,favorablePropagation2}. 

From \cite[Eq. (18)--(19)]{favorablePropagation2}, we know that for the case of unlimited physical array space, 
\begin{align}
E \bigg\{\frac{1}{N}\mathbf{h}_k^H \mathbf{h}_j\bigg\} &\rightarrow 0 \label{Ngo} \\
{\rm var} \bigg\{\frac{1}{N}\mathbf{h}_k^H \mathbf{h}_j\bigg\} &= \frac{1}{N}-\frac{1}{N^2}. \label{Ngo2}
\end{align}
Comparing \eqref{asy_mom1} and \eqref{asy_mom2} with \eqref{Ngo}-\eqref{Ngo2}, we can infer that the corresponding expressions coincide if $d_0=\frac{N}{2}$, in which case both moments will converge smoothly to zero. On the contrary, if the physical array space is fixed (i.e. it does not grow with $N$), the mean and variance of the inner product, which indicate the level of inter-user interference, will converge to non-zero constant values that are inversely proportional to $d_0$.

\section{Numerical Results}
\begin{figure} [tb]
\begin{center}
\includegraphics[width=\fsize\textwidth, keepaspectratio]
{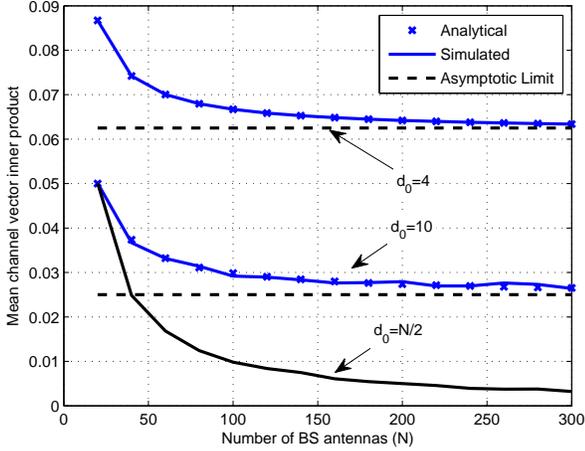} \caption{Mean channel vector inner product $E \left\{\frac{1}{N}\mathbf{h}_k^H \mathbf{h}_j \right\}$ vs. $N$, for massive MIMO with limited/unlimited physical space.} \label{innerprodN}
\end{center}
\end{figure}

\begin{figure} [tb]
\begin{center}
\includegraphics[width=\fsize\textwidth, keepaspectratio]
{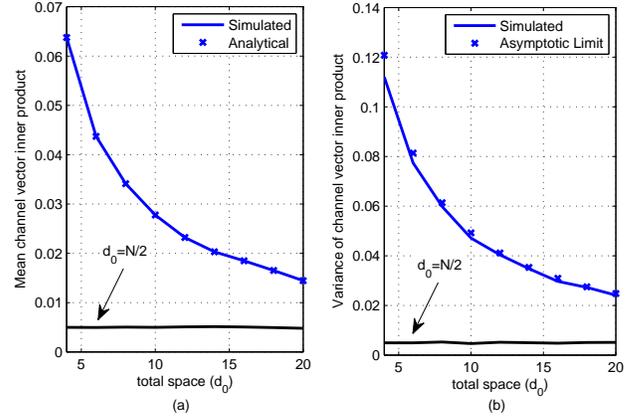} \caption{ a) Mean channel vector inner product $E \left\{\frac{1}{N}\mathbf{h}_k^H \mathbf{h}_j \right\}$ vs. $d_0$ and b) variance of channel vector inner product ${\rm var} \left\{\frac{1}{N}\mathbf{h}_k^H \mathbf{h}_j \right\}$ vs. $d_0$, for massive MIMO with limited/unlimited physical space ($N=200$).\color{black}} \label{innerprodd0}
\end{center}
\end{figure}


This section illustrates the analytical performance and results of Monte-Carlo simulations for the massive MIMO scenario under study in comparison with the case of massive MIMO with unlimited physical space. The simulations assume a linear antenna array at the BS, $K=10$ users and the steering vector channel model of \eqref{LOS}. 

Figure \ref{innerprodN} shows the simulated and analytical mean inner product \eqref{exact_mom1} of the channel vectors for an increasing number of antennas $N$, along with the theoretical limit of \eqref{asy_mom1}. The cases for $d_0=4$ and $d_0=10$ are shown and it can be seen that \eqref{exact_mom1} yields an exact match to simulation. More importantly, it is evident that, unlike the case of unlimited physical space, for the space-constrained deployment the inner product does not converge to zero, but to $\frac{1}{4d_0}$ as shown in our analysis. This implies that interference is not completely nulled, which makes MRT precoding non-optimal for this scenario.

In Fig. \ref{innerprodd0}(a), the simulated and analytical mean inner product of the channel vectors is shown for an increasing total ULA space $d_0$. The mean inner product is also shown for the ULA deployment without space constraints for reference. A non-negligible deviation between the two models can be seen, which becomes significant as the total space $d_0$ decreases.

In Fig. \ref{innerprodd0}(b), the simulated and asymptotic variance of the inner product of the channel vectors is shown for am increasing total ULA space $d_0$. The analytical asymptotic limit illustrates expression \eqref{asy_mom2}, which yields excellent tightness with the simulation results even for small $d_0$.  Again, the case of ULA deployment without space constraints is shown for reference and a deviation between the two models can be seen.

\begin{figure} [tb]
\begin{center}
\includegraphics[width=\fsize\textwidth, keepaspectratio]
{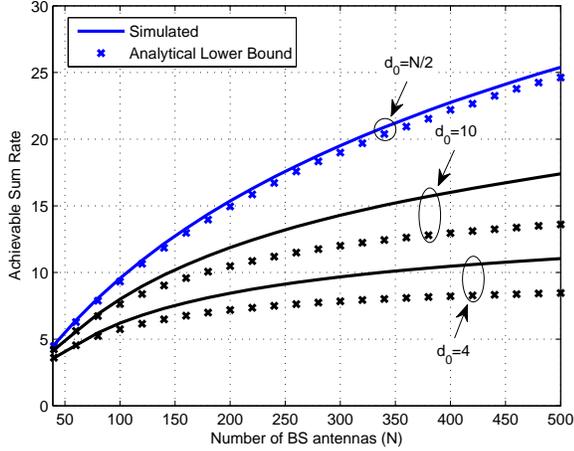} \caption{Achievable sum rate for massive MIMO with limited/unlimited  physical space.} \label{sumrateN}
\end{center}
\end{figure}

Finally, Fig. \ref{sumrateN} illustrates the effect of the above observations on the achievable sum rate of the system under MRT, for a transmit SNR of $\rho=10$dB where we have accounted for additional shadowing and propagation losses of 30dB. The Jensen's lower bound in \eqref{rate3} can be easily evaluated in closed-form using the results of Proposition 1; unfortunately, $R_L$ is relatively loose for the channel model considered herein, since ${\rm var}\left\{\frac{1}{N^2}|\mathbf{h}_k^H \mathbf{h}_j|^2\right\}$ is much higher compared, for example, to the Rayleigh fading case considered in \cite{favorablePropagation}. The achievable sum rates are shown for the cases $d_0=4, 10$ and compared to the deployment over unlimited physical space. Contrary to the latter case, a saturation in the sum rates can be observed for the space-constrained deployment, where, as expected,
the ceiling of the achievable sum rate starts diminishing with increasing total space $d_0$.

\section*{Conclusion}
The favorable propagation characteristics of massive MIMO have been examined under space-constrained antenna deployments. Contrary to current understanding, it was shown that when increasing the numbers of antennas that are deployed in a fixed total space, the inner product of the channel vectors tends to a non-zero value. This phenomenon implies that inter-user interference does not vanish in the massive MIMO regime, thereby making MRT non-optimal. The above discussion poses a fundamental divergence from existing intuition and motivates a further study of the resulting space-constrained deployments and the development of suitable transmission schemes.

\appendices
\section{Useful Lemma}
\begin{lemma}
Defining ${\rm sinc}(x)\triangleq \sin(x)/x$ and for $ 0\leq \beta \leq \pi $, we have that
\begin{align} \label{EI2}
\sum_{m=0}^{\infty} {\rm sinc}^2\left(\beta m\right)&=\frac{1}{2} \left(1+\frac{\pi}{\beta}\right).
\end{align}
\end{lemma}
\begin{IEEEproof}
Using standard trigonometric properties, the infinite summation in (\ref{EI2}) can be expanded as follows:
\begin{align}\label{EI2a}
\sum_{m=0}^{\infty} {\rm sinc}^2\left(\beta m\right) =1 + \frac{1}{2 \beta^2}\sum_{m=1}^{\infty}\frac{1}{m^2}-\sum_{m=1}^{\infty}\frac{\cos(2\beta m)}{2\beta^2m^2}.
\end{align}
The first infinite sum in (\ref{EI2a}) converges to $\pi^2/6$ while the second one can be evaluated with the aid of 
\cite[Eq. (5.4.2.12)]{Prudnikov}. Putting everything together, yields the desired result.
\end{IEEEproof}

\section{Proof of Proposition \ref{Asy_prop}} \label{app:LARGE}
{ The proof of \eqref{asy_mom1} follows trivially by recalling \eqref{exact_mom1} and Lemma 1, where the condition on $a$ holds for large $N$. 
Regarding the square inner product $E \left\{\left|\mathbf{h}_k^H \mathbf{h}_j \right| ^2\right\}$, this can be expanded as follows:}
\begin{align}\label{sum1}
&E \left\{\left|\mathbf{h}_k^H \mathbf{h}_j \right| ^2\right\} =\sum_{m_1=0}^{N-1}\sum_{m_2=0}^{N-1}{\rm sinc}^2\left(a(m_1-m_2)\right) \nonumber\\
&{ =N +\sum_{m_1=0}^{N-1}\sum_{\scriptsize\begin{array}{c}m_2=0\\ m_2\neq m_1\end{array}}^{N-1}\frac{1-\cos(2a(m_1-m_2))}{2a^2(m_1-m_2)^2}}\nonumber\\
&= N +\frac{ 1}{2a^2}\sum_{m_1=0}^{N-1}\sum_{\scriptsize\begin{array}{c}m_2=0\\ m_2\neq m_1\end{array}}^{N-1}\frac{ 1}{(m_1-m_2)^2}\nonumber\\
  &-\frac{ 1}{2a^2}\sum_{m_1=0}^{N-1}\sum_{\scriptsize\begin{array}{c}m_2=0\\ m_2\neq m_1\end{array}}^{N-1} \frac{ \cos(2a(m_1-m_2))}{(m_1-m_2)^2}. \end{align}
After some tedious but straightforward manipulations, it can be shown that, for $N \rightarrow \infty$, the first sum in \eqref{sum1} converges to 
\begin{align}\label{sum2}
\sum_{m_1=0}^{N-1}\sum_{\scriptsize\begin{array}{c}m_2=0\\ m_2\neq m_1\end{array}}^{N-1}\frac{ 1}{(m_1-m_2)^2}\rightarrow  2N\frac{\pi^2}{6}-\sum_{k=1}^{\infty}\frac{1}{k}.
\end{align}
Likewise, the second sum in \eqref{sum1} converges to
\begin{align}\label{sum3}
\sum_{m_1=0}^{N-1}\sum_{\scriptsize\begin{array}{c}m_2=0\\ m_2\neq m_1\end{array}}^{N-1} \frac{ \cos(2a(m_1-m_2))}{(m_1-m_2)^2} & \nonumber\\
\rightarrow 2N\left(a^2-\pi a+\frac{\pi^2}{6}\right)
&-\sum_{k=1}^{\infty}\frac{\cos(2ka)}{k}.
\end{align}
{Combining \eqref{sum1}--\eqref{sum3} and simplifying, yields the desired result. Finally, the asymptotic result for the variance of the inner product is a direct combination of \eqref{asy_mom1}, \eqref{asy_sqp} with \eqref{Var_def}.}

\end{document}